\documentclass[aps,prl,twocolumn,groupedaddress,showpacs,amsmath,amssymb]{revtex4-1}
\usepackage[dvips]{graphicx,color}
\usepackage{booktabs}
\usepackage{tensor}
\begin{document}

\title{Gravitational Field Tensor}

\author{Stephen M. Barnett}

\email{stephen.barnett@glasgow.ac.uk}

\affiliation{School of Physics and Astronomy, University of Glasgow, Glasgow G12 8QQ, UK}

\date{\today}

\begin{abstract}
We present a tensorial relative of the familiar affine connection and argue that it should be regarded as the
gravitational field tensor.  Remarkably, the Lagrangian density  expressed in terms of this tensor has a simple form, 
which depends only on the metric and its first derivatives and, moreover, is a true scalar quantity.  The geodesic
equation, moreover, shows that our tensor plays a role that is strongly reminiscent of the gravitational field in Newtonian
mechanics and this, together with other evidence, which we present, leads us to identify it as the gravitational field
tensor.  We calculate the gravitational field tensor for the Schwarzschild metric.  We suggest some of the advantages 
to be gained from applying our tensor to the study of gravitational waves.
\end{abstract}

\maketitle

That physical quantities should be represented as tensors is one of the fundamental ideas in the general theory of relativity.
It is this way that we arrive naturally at a theory that applies in any coordinate system and for any state of motion.  It comes as
something of a surprise, therefore, to discover in one's first course on the subject that the affine connection
$\tensor{\Gamma}{^\lambda_{\mu\nu}}$, or Christoffel symbol of the second kind, is \emph{not} a tensor.  
Despite this shortcoming, it appears naturally in the
geodesic equation for the motion of a particle playing a role akin to the electromagnetic field tensor, 
$\tensor{F}{^{\mu\nu}}$, in the equation of motion for a charged particle.  In this sense, at least, the connection is a 
gravitational analogue of the electromagnetic field.  The geodesic equation illustrates clearly, however, that the affine
connection cannot be a tensor, for were it to be a tensor then the existence of a local inertial frame, in which the affine
connection vanishes would necessarily require the tensor itself to be zero in all coordinate systems.

It is well-known that although the affine connection is not a tensor, the \emph{difference} between any two affine connections 
\emph{is} a tensor \cite{Eisenhardt}.
Consider the tensor quantity defined to be the difference between the two connections, $\tensor{\Gamma}{^\lambda_{\mu\nu}}$
and $\tensor{\tilde\Gamma}{^\lambda_{\mu\nu}}$, expressed in terms of a common set of coordinates:
\begin{equation}
\label{DeltaDef}
\tensor{\Delta}{^\lambda_{\mu\nu}} = \tensor{\Gamma}{^\lambda_{\mu\nu}} - \tensor{\tilde\Gamma}{^\lambda_{\mu\nu}} \, .
\end{equation}
Each connection is derived from a metric, $\tensor{g}{_{\mu\nu}}$, $\tensor{\tilde g}{_{\mu\nu}}$, in the usual way \cite{Diracbook}.
We choose $\tensor{g}{_{\mu\nu}}$ to be the usual metric tensor for our space--time, and thus $\tensor{\Gamma}{^\lambda_{\mu\nu}}$
is the true connection.  We leave unspecified, for the present, the precise form of $\tensor{\tilde g}{_{\mu\nu}}$ and thus of 
$\tensor{\tilde\Gamma}{^\lambda_{\mu\nu}}$, other than to state that we shall choose it to correspond to a metric for a flat 
space so that the Ricci tensor vanishes ($\tensor{\tilde R}{_{\mu\nu}}=0$) \cite{footnote}.  
Our task will be to investigate the role of $\tensor{\Delta}{^\lambda_{\mu\nu}}$ within the general theory of relativity.  We work
throughout with the natural system of units in which Newton's gravitational constant, $G$, and the speed of light, $c$, are
both set to unity.

We note, first, that the difference between the Riemann curvature and that for a space--time with connection 
$\tensor{\tilde\Gamma}{^\lambda_{\mu\nu}}$ has a natural and simple expression in terms of 
$\tensor{\Delta}{^\lambda_{\mu\nu}}$ \cite{Eisenhardt}:
\begin{eqnarray}
\label{RiemannDiff}
\tensor{R}{^\beta_{\nu\rho\sigma}} - \tensor{\tilde R}{^\beta_{\nu\rho\sigma}} &=&
\tensor{\Delta}{^\beta_{\nu\sigma ;\rho}} - \tensor{\Delta}{^\beta_{\nu\rho ;\sigma}}  \nonumber \\
& & + \tensor{\Delta}{^\alpha_{\nu\sigma}}\tensor{\Delta}{^\beta_{\alpha\rho}}
- \tensor{\Delta}{^\alpha_{\nu\rho}}\tensor{\Delta}{^\beta_{\alpha\sigma}} \, ,
\end{eqnarray}
where ; denotes covariant differentiation and we follow, in this manuscript, the conventions adopted by Dirac \cite{Diracbook}.
This formula is reminiscent of the expression for the Riemann tensor written in terms of derivatives and products of 
the connection but has the important feature that each term in it is a tensor.  This follows directly from the fact that 
$\tensor{\Delta}{^\lambda_{\mu\nu}}$ is a tensor.  A correspondingly simple expression for the difference in the Ricci 
tensor follows directly from that for the difference in the Riemann tensor:
\begin{eqnarray}
\label{RicciDiff}
\tensor{R}{_{\nu\rho}} - \tensor{\tilde R}{_{\nu\rho}} &=&
\tensor{\Delta}{^\beta_{\nu\beta ;\rho}} - \tensor{\Delta}{^\beta_{\nu\rho ;\beta}}  \nonumber \\
& & + \tensor{\Delta}{^\alpha_{\nu\beta}}\tensor{\Delta}{^\beta_{\alpha\rho}}
- \tensor{\Delta}{^\alpha_{\nu\rho}}\tensor{\Delta}{^\beta_{\alpha\beta}} \, .
\end{eqnarray}
As we have selected $\tensor{\tilde\Gamma}{^\lambda_{\mu\nu}}$ such that $\tensor{\tilde R}{_{\mu\nu}}=0$, this is also 
an expression for the Ricci tensor, $\tensor{R}{_{\nu\rho}}$, alone.  A further contraction using the inverse of the metric,
$\tensor{g}{^{\nu\rho}}$, gives, naturally enough, the curvature scalar.

As a first application of the above ideas, we consider the action for the gravitational field, which is usually given in
the Einstein-Hilbert form \cite{Diracbook,Maggiore,Keifer}:
\begin{equation}
I_{EH} = \frac{1}{16\pi} \int d^4x \sqrt{-g} \: R \, ,
\end{equation}
where $R = \tensor{g}{^{\mu\nu}}\tensor{R}{_{\mu\nu}}$ is the curvature scalar.  Variation of this action leads, of course,
to the free-field Einstein equation, but the task of extracting this is complicated by the fact that the Lagrangian density
depends on the metric, its first derivatives and also \emph{second} derivatives.  It is possible to use integration by parts 
to remove the second derivatives and thereby simplify the derivation.
This widely adopted procedure leads to the Lagrangian density \cite{Diracbook,Schrodinger,Feynman,LL,Hobson}:
\begin{equation}
\label{Diracdensity}
\mathcal{L}_D = 
\frac{1}{16\pi}\tensor{g}{^{\mu\nu}}
\left(\tensor{\Gamma}{^\sigma_{\mu\nu}}\tensor{\Gamma}{^\rho_{\sigma\rho}}
- \tensor{\Gamma}{^\rho_{\mu\sigma}}\tensor{\Gamma}{^\sigma_{\nu\rho}}\right) \, .
\end{equation}
This is \emph{not} a scalar quality, however, by virtue of the fact that the affine connection is not a tensor 
\cite{Diracbook,Schrodinger,Hobson} but was, nevertheless, used as the staring point in Dirac's Hamiltonian formulation
of gravitation \cite{Dirac58}.  If we use (\ref{RicciDiff}) in place of the familiar expression for the Ricci tensor, however,
then a similar integration by parts leads to the Lagrangian density \cite{footnote1}:
\begin{equation}
\label{LagrangianDelta}
\mathcal{L}_\Delta = \frac{1}{16\pi}\tensor{g}{^{\mu\nu}}
\left(\tensor{\Delta}{^\alpha_{\lambda\alpha}}\tensor{\Delta}{^\lambda_{\mu\nu}}
- \tensor{\Delta}{^\lambda_{\alpha\mu}}\tensor{\Delta}{^\alpha_{\lambda\nu}}\right) \, .
\end{equation}
That this is a scalar follows directly from the fact that $\tensor{\Delta}{^\lambda_{\mu\nu}}$ is a tensor.  It is also
clear that this form of the Lagrangian density depends only on the metric and its first derivatives.  The highly 
desirable situation of having only first order derivatives, and therefore straightforward Euler-Lagrange equations,
has motivated the search for alternative Lagrangians, yet none of the existing rivals we are aware of have the 
simplicity of $\mathcal{L}_\Delta$ \cite{Palatini,Lanczos62}.  We note, in particular, that it is bilinear in the tensor 
$\tensor{\Delta}{^\lambda_{\mu\nu}}$, much as the electromagnetic Lagrangian density is bilinear in the electromagnetic
field tensor.  There is a long history of attempts to formulate an action principle on the basis of bilinear combinations of 
the \emph{curvature}, as opposed to the linear form embodied in the Einstein-Hilbert action \cite{Weyl,Lanczos38}, and 
also to express other physically significant properties in this way \cite{Bel,Robinson,Misner,Belinski}.  Our form for the
Lagrangian density, however, emphasises the role of the tensor $\tensor{\Delta}{^\lambda_{\mu\nu}}$ rather than the 
curvature.  We elaborate on this point further below, but consider first an example of the explicit the form of our tensor 
and of its application to the study of gravitational waves.

Working with $\tensor{\Delta}{^\lambda_{\mu\nu}}$ rather than the connection or the metric itself may present
significant calculational advantage, by making a suitable choice of flat-space metric $\tensor{\tilde g}{_{\mu\nu}}$.
As a simple illustration of this we consider the Schwarzschild metric corresponding to a mass $m$ localised at
the origin, with the tilde space corresponding to the absence of this mass.  If we employ the natural spherical polar 
coordinates then there are 9 non-vanishing distinct components of the connection \cite{Diracbook} but only 5 non-zero distinct components of $\tensor{\Delta}{^\lambda_{\mu\nu}}$:
\begin{eqnarray}
\label{Schwarzschild}
\tensor{\Delta}{^1_{00}} &=& \frac{m}{r^2}\left(1 - \frac{2m}{r}\right)  \nonumber \\
\tensor{\Delta}{^1_{11}} &=& -\frac{m}{r^2}\left(1 - \frac{2m}{r}\right)^{-1}  \nonumber \\
\tensor{\Delta}{^1_{22}} &=&  2m  \nonumber \\
\tensor{\Delta}{^1_{33}} &=& 2m\sin^2\theta \nonumber \\
\tensor{\Delta}{^0_{10}} &=& \frac{m}{r^2}\left(1 - \frac{2m}{r}\right)^{-1} \, ,
\end{eqnarray}
where the coordinates are numbered as $(0, 1, 2, 3) = (t, r, \theta , \phi)$.  The corresponding Ricci tensor is
zero (away from the origin) as is the curvature scalar.  Proving this using our tensor is far simpler than using the 
affine connection as $16\pi \mathcal{L}_\Delta$, $\tensor{\Delta}{^\lambda_{\mu\lambda}}$ and
$\tensor{\Delta}{^\lambda_{\mu\nu ; \lambda}}\tensor{g}{^{\mu\nu}}$, which combine to give $R$, are each 
\emph{separately} zero.

The fact that the tensor $\tensor{\Delta}{^\lambda_{\mu\nu}}$ is the difference between connections suggests, in particular,
application to situations in which it is appropriate to \emph{linearize} about a background metric and hence to the
theory of gravitational waves \cite{Maggiore,Schutz}.  The linearization usually leads to a wave equation for the small deviation
of the metric associated with the wave, but its derivation requires the choice of a suitable gauge choice (the transverse
traceless gauge) in a particular coordinate system \cite{Schutz}.  Gravitational waves propagating in a region of free space
satisfy $R_{\mu\nu} = 0$.  Use of the tensors $\tensor{\Delta}{^\lambda_{\mu\nu}}$ and working to first order then leads 
directly to the equation 
\begin{equation}
\label{Deltawaveequation}
\tensor{\Delta}{^\alpha_{\mu\alpha ; \nu}} - \tensor{\Delta}{^\alpha_{\mu\nu ; \alpha}} = 0 \, .
\end{equation}
If we rewrite this in terms of the small correction to the metric in a Minkowski background and choose the transverse
traceless gauge then this becomes the familiar gravitational wave-equation.  It should be emphasised, however, that this
equation is a \emph{tensor} equation and hence applies in any coordinate system.  The tensor 
$\tensor{\Delta}{^\lambda_{\mu\nu}}$ is, moreover, gauge-invariant at least to this lowest order.  To see this we need only note that
a gauge transformation corresponds to a local coordinate transformation \cite{Kibble} and that we must, therefore transform
\emph{both} $\tensor{g}{_{\mu\nu}}$ and $\tensor{\tilde g}{_{\mu\nu}}$:
\begin{eqnarray}
\label{Gauge}
\tensor{g}{_{\mu\nu}} &\rightarrow & \tensor{g}{_{\mu\nu}} - \xi_{\mu,\nu} - \xi_{\nu,\mu} \nonumber \\
\tensor{\tilde g}{_{\mu\nu}} &\rightarrow & \tensor{\tilde g}{_{\mu\nu}} - \xi_{\mu,\nu} - \xi_{\nu,\mu}
\end{eqnarray}
where the , denotes differentiation as usual \cite{Diracbook}.  Hence our gravitational wave equation (\ref{Deltawaveequation})
is both a tensor equation, and hence holds in any coordinate system, and it is also gauge-invariant.  In this sense it should
be viewed as the natural analogue of the free-field Maxwell equations for electromagnetic waves \cite{MaxGrav}.  The symmetries
and conservation laws for gravitational waves may be lifted from the symmetries of the background metric, 
$\tensor{\tilde g}{_{\mu\nu}}$, by applying Noether's theorem \cite{Neuenschwander} to the action
\begin{equation}
\label{NoetherWaves}
I = \frac{1}{16\pi}\int d^4x \sqrt{-\tilde g} \: \tensor{\tilde g}{^{\mu\nu}}
\left(\tensor{\Delta}{^\alpha_{\lambda\alpha}}\tensor{\Delta}{^\lambda_{\mu\nu}}
- \tensor{\Delta}{^\lambda_{\alpha\mu}}\tensor{\Delta}{^\alpha_{\lambda\nu}}\right) \, ,
\end{equation}
where the tensors $\tensor{\Delta}{^\lambda_{\mu\nu}}$ are restricted to first order in the difference between 
$\tensor{\tilde g}{_{\mu\nu}}$ and $\tensor{g}{_{\mu\nu}}$.

It remains to make explicit the case for referring to the tensor $\tensor{\Delta}{^\lambda_{\mu\nu}}$ as
the gravitational field tensor.  There are three strong indications of this and the combination of these is compelling.
These are (i) the form of the geodesic equation for the motion of a test particle, (ii) comparison with corresponding
quantities in electromagnetism and (iii) the existence of an analogy with Yang-Mills theories.  Let us take each of
these in turn.

\emph{(i) The geodesic equation}.   The geodesic equation for the motion of a test particle is \cite{Diracbook}
\begin{eqnarray}
\label{geodesic}
\frac{d^2 x^\lambda}{d \tau^2} &=& - \tensor{\Gamma}{^\lambda_{\mu\nu}}\frac{d x^\mu}{d\tau}\frac{dx^\nu}{d\tau} 
\nonumber \\
&=& - \tensor{\Delta}{^\lambda_{\mu\nu}}\frac{d x^\mu}{d\tau}\frac{dx^\nu}{d\tau}
 - \tensor{\tilde \Gamma}{^\lambda_{\mu\nu}}\frac{d x^\mu}{d\tau}\frac{dx^\nu}{d\tau} \, .
\end{eqnarray}
In this equation of motion the term containing $\tensor{\tilde \Gamma}{^\lambda_{\mu\nu}}$ describes the motion
as it would be in the \emph{absence} of the body or bodies responsible for the curvature.  The term containing the
tensor $\tensor{\Delta}{^\lambda_{\mu\nu}}$ provides the modification of this motion due to the presence of
the gravitating bodies, that is the gravitational field.  In this sense it provides the natural analog of the Newtonian 
gravitational force, as encapsulated in Newton's first law of motion \cite{Chandrasekhar}, in that it induces the 
deviation from the uniform motion associated with $\tensor{\tilde \Gamma}{^\lambda_{\mu\nu}}$ acting alone.

\emph{(ii) Comparison with electromagnetism}.  Analogies between electromagnetism and gravitation have often
been applied as an aid to understanding and teaching phenomena within general relativity \cite{Hobson,MaxGrav}.
This analogy enhances the case for identifying $\tensor{\Delta}{^\lambda_{\mu\nu}}$ as the gravitational field tensor.
To see this we recall that in electromagnetism we introduce the four-potential $A^\mu$ from which we can form the
field $F^{\mu\nu}$ by differentiation.  Finally the fields are coupled to charged matter through Maxwell's equations
in which the derivatives of the fields appear.  All of these quantities, the four-potential, the field and its derivatives are
tensor quantities \cite{footnote2}.  

A strongly analogous scheme for gravitational fields treats the metric,
$g_{\mu\nu}$, as a potential.  From this we obtain the connection, $\tensor{\Gamma}{^\lambda_{\mu\nu}}$, by
differentiation and thence, via further differentiation, the Riemann tensor and the equation of motion expressed
in terms of the Ricci tensor.  As with electromagnetism, each of these are tensors with the \emph{exception} of
the connection.  But for this, it would be natural in this scheme to associate the connection with the gravitational field, 
in analogy with the electromagnetic field tensor, $F^{\mu\nu}$.  We can complete the analogy with electromagnetism 
by replacing the affine connection with the gravitational field tensor $\tensor{\Delta}{^\lambda_{\mu\nu}}$.  Like the
connection, it is obtained from the metric by differentiation and further differentiation of it leads to the Riemann
tensor and to the Ricci tensor, which is coupled to matter sources in the Einstein field equation.  The relationships
between these quantities are depicted in the tables.  

In table \ref{table_a} we present the electromagnetic potential 
and field together with the governing Maxwell equation that couples the fields to the material sources.  In the second 
line we have the analogous quantities for gravity.  Note the appearance of the connection, which is the only 
quantity in the table that is \emph{not} a tensor.  In table \ref{table_b} the connection is replaced by our gravitational
field tensor, so that every quantity in the table is a tensor.

\begin{table}[!h]
\begin{center}
\begin{tabular}{@{}ccc@{}}
\toprule
\hline
Potential & \quad Field? & \quad Field Equation \\
\hline \hline
\midrule
$\tensor{A}{^\mu}$ & $\tensor{F}{^{\mu\nu}}$ & $\tensor{F}{^{\mu\nu}_{;\nu}}=\tensor{j}{^\mu}$ \\
$\tensor{g}{^{\mu\nu}}$ &
\color{red}$\tensor{\Gamma}{^{\lambda}}{_{\mu\nu}}$ &
\quad $\tensor{G}{_{\mu\nu}}=-8\pi\tensor{T}{_{\mu\nu}}$
\\
\hline
\bottomrule
\end{tabular}
\caption{The familiar quantities in electromagnetism and in gravitation.  In both cases we progress
from potential to field to field equation by differentiation.} \label{table_a}
\end{center}
\end{table}

\begin{table}[!h]
\begin{center}
\begin{tabular}{@{}ccc@{}}
\hline
\toprule
Potential & \quad Field Tensor & \quad Field Equation \\
\hline \hline
\midrule
$\tensor{A}{^\mu}$ & $\tensor{F}{^{\mu\nu}}$ & $\tensor{F}{^{\mu\nu}_{;\nu}}=\tensor{j}{^\mu}$ \\
$\tensor{g}{^{\mu\nu}}$ &
\color{green}$\tensor{\Delta}{^{\lambda}}{_{\mu\nu}}$ &
\quad $\tensor{G}{_{\mu\nu}}=-8\pi\tensor{T}{_{\mu\nu}}$
\\
\hline
\bottomrule
\end{tabular}
\caption{A more natural assignment in which the non-tensorial connection is replaced by the
tensor $\tensor{\Delta}{^{\lambda}}{_{\mu\nu}}$.} \label{table_b}
\end{center}
\end{table}

\emph{(iii)  Analogy with other field theories}. Finally, and most speculatively, we note that the Lagrangian density 
when expressed in terms of the tensor $\tensor{\Delta}{^\lambda_{\mu\nu}}$ is bilinear.  In this sense it is
reminiscent of the Lagrangian density for Yang-Mills theories and electromagnetism \cite{YM,Weinberg}:
\begin{equation}
\label{YMLagrangian}
\mathcal{L}_{YM} = -\frac{1}{4}{\bf F}^{\alpha\beta}\cdot{\bf F}_{\alpha\beta} \, .
\end{equation}
Although our Lagrangian density, $\mathcal{L}_\Delta$, is not explicitly of this form, it is bilinear in the gravitational 
field tensor.  The formal similarity with Yang-Mills theories can be made yet stronger if we add to Eq. (\ref{YMLagrangian})
the zero-valued quantity $\frac{1}{4}\tensor{\bf F}{^\lambda_\lambda}\cdot\tensor{\bf F}{^\mu_\mu}$.  It is possible that this 
analogy (or similarity in form) between $\mathcal{L}_\Delta$ and the Yang-Mills Lagrangian density might suggest new 
directions in the study of quantum effects in gravity.

\begin{acknowledgments}
I am grateful to John Cameron, Rob Cameron, Jim Cresser, Sarah Croke, Claire Gilson, Norman Gray, Martin Hendry,
Fiona Speirits, Graham Woan and Alison Yao for helpful comments and suggestions.
\end{acknowledgments}


\begin{thebibliography}{10}
\bibitem{footnote}
The equation $\tensor{\tilde R}{_{\mu\nu}}=0$ provides 10 independent conditions that we can use to fix
the $\tensor{\tilde g}{_{\mu\nu}}$, up to a gauge choice, and hence the connection 
$\tensor{\tilde \Gamma}{^\lambda_{\mu\nu}}$.
\bibitem{footnote1}
This follows directly from the fact that $\sqrt{-g}\tensor{\Delta}{^\beta_{\nu\beta;\rho}}\tensor{g}{^{\nu\rho}}=
\left(\sqrt{-g}\tensor{\Delta}{^\beta_{\nu\beta}}\tensor{g}{^{\nu\rho}}\right)_{,\rho}$ (with a similar relation 
for $\tensor{\Delta}{^\beta_{\nu\rho;\beta}}$) so that application of Gauss's theorem to the action integral turn this 
into a surface term that may be taken to vanish at a large distance.
\bibitem{footnote2}
This is true if we use the appropriate form of Maxwell's equations containing covariant derivatives \cite{LL}.
\end{thebibliography}
\end{document}